\documentclass[runningheads]{llncs}

\usepackage[T1]{fontenc}
\usepackage{listings}
\usepackage{float}
\usepackage{graphicx}
\usepackage{listings}
\usepackage{booktabs}
\usepackage{multirow}
\usepackage{alltt}
\usepackage{subcaption}
\usepackage{xspace}
\usepackage{paralist}
\usepackage{comment}
\usepackage{amsfonts}
\usepackage{pdfpages}
\usepackage{color,soul}
\usepackage{xcolor}
\usepackage{fp}
\usepackage[english]{babel}
\usepackage{todonotes}
\newcommand{\ie}{i.e.,\xspace}
\newcommand{\eg}{e.g.,\xspace}

\usepackage{tikz}
\usepackage{circledsteps}
\definecolor{mycolor}{RGB}{255,241,79}

\definecolor{mycolorB}{RGB}{79,241,255}
\newcommand*\rectangled[1]{\tikz[baseline=(char.base)]{\node[shape=rectangle,fill=mycolorB,draw,inner sep=1pt] (char) {\normalfont#1};}}

\usepackage{hyperref}

\begin{document}
\title{Post-hoc LLM-Supported Debugging of Distributed Processes}
\author{Dennis Schiese\inst{1} \and
Andreas Both\inst{1}}
\authorrunning{Dennis Schiese, Andreas Both}
\institute{\includegraphics[height=2ex]{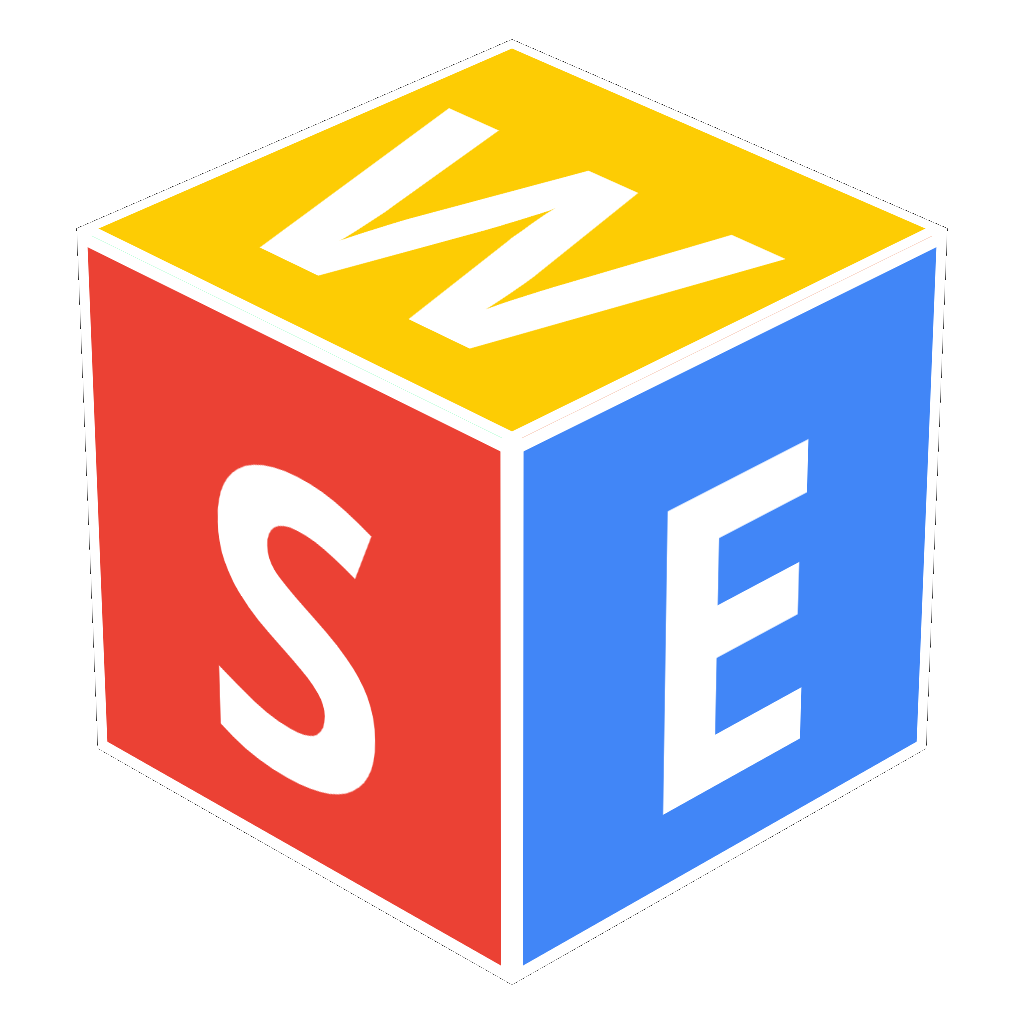} Web \& Software Engineering Research Group\\Leipzig University of Applied Sciences}
\maketitle              
\begin{abstract}

In this paper, we address the problem of manual debugging, which nowadays remains resource-intensive and in some parts archaic. This problem is especially evident in increasingly complex and distributed software systems. Therefore, our objective of this work is to introduce an approach that can possibly be applied to any system, at both the macro- and micro-level, to ease this debugging process. This approach utilizes a system's process data, in conjunction with generative AI, to generate natural-language explanations. These explanations are generated from the actual process data, interface information, and documentation to guide the developers more efficiently to understand the behavior and possible errors of a process and its sub-processes. Here, we present a demonstrator that employs this approach on a component-based Java system. 
However, our approach is language-agnostic. Ideally, the generated explanations will provide a good understanding of the process, even if developers are not familiar with all the details of the considered system.
Our demonstrator is provided as an open-source web application that is freely accessible to all users.

\end{abstract}
%
%
%

\section{Introduction}

\begin{figure}[t]
    \fbox{
    \centering
    \includegraphics[width=.95\textwidth]{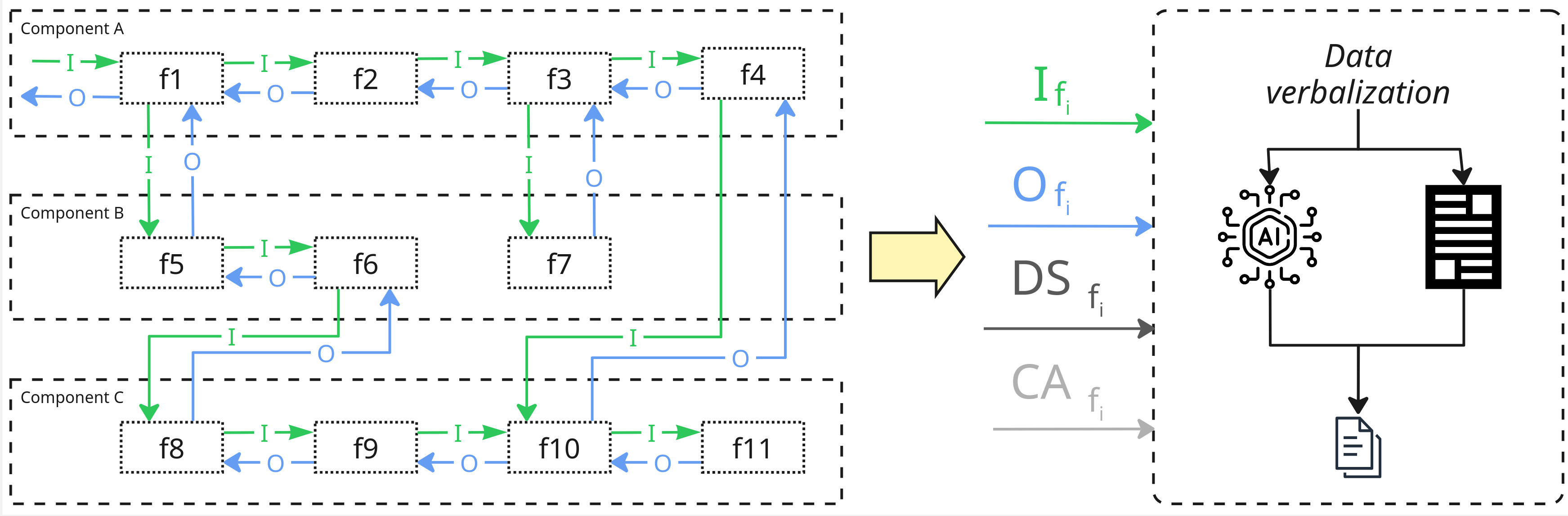}}
\vspace{-1.5ex}    \caption{An abstract process involving three interacting components. Each unit (\textit{f}) may call another with or without arguments and possibly receive a return value. These calls, along with values like input ($I_{f_i}$), output ($O_{f_i}$), docstring ($DS_{f_i}$), and caller ($CA_{f_i}$), are logged and used by a \textit{Data verbalizer} to generate explanations for $f_i$, using templates or generative AI. 
    }
    \label{fig:process}
    \vspace{-3.5ex}
\end{figure}

Software development, as with many other areas, is subject to rapid growing automation, which has a considerable effect on the efficiency with which software is developed. 
As a result, the volume and complexity of existing software continue to grow. This is particularly evident in the context of distributed systems, where processes utilize multiple software components (\eg Web services), and in the development of complex software architectures, where intricate dependencies and large codebases add further challenges.
However, as software development accelerates, traditional workflows such as conception, implementation, debugging, or testing remain fundamental, even as AI agents begin to adopt them. 
In the case of debugging, concepts like \textit{automated program repair} (APR) offer many different techniques and tools to tackle this problem \cite{dikici_advancements_2025,10.1145/3631974}. 
Despite their increasing performance, these approaches encounter instances of failure. 
In such cases, manual debugging becomes a necessary process.

In this demonstration, one of the considered processes -- consisting of just two components (Web services) -- can trigger over 5000 trackable inter- and intra-component method calls (excluding external libraries). 
The manual debugging of such a process is cumbersome and time-consuming. 
To address this challenge, we proposed in \cite{schiese2024semantics,schiese2024icwi} an approach that inspects a system during runtime and verbalizes process' data post-hoc for each method. 
Here, we extended this approach with a clear focus on component-oriented explanations (see Fig.~\ref{fig:process}) to ease the debugging process and implemented a corresponding demonstrator.
In order to reduce the overhead of having to search through too many explanations, Large Language Models (LLMs) are employed in order to extract the most relevant aspects of the sub-called methods' explanations and to integrate them into the natural-language explanation of the current method. 
We aim to propagate all relevant aspects of all subprocesses, while less relevant aspects may not be used for parent methods, s.t., developers receive a compact explanation of what actually happened within the specific process.
This or analogous approaches, \ie any work that utilizes process data to generate explanations with the usage of generative AI for manual debugging, are, to the best of our knowledge, not currently known. 
However, in the field of APR, the utilization of generative AI has been found to be a prevailing trend \cite{10376418,10172803}; 
\cite{Kang2024} proposed an approach towards (automated) debugging while aiming for LLM-supported patch generation.

\section{LLM-driven Debugging for Component-based Systems}

\begin{figure}[t]
    \centering
    \includegraphics[width=\textwidth]{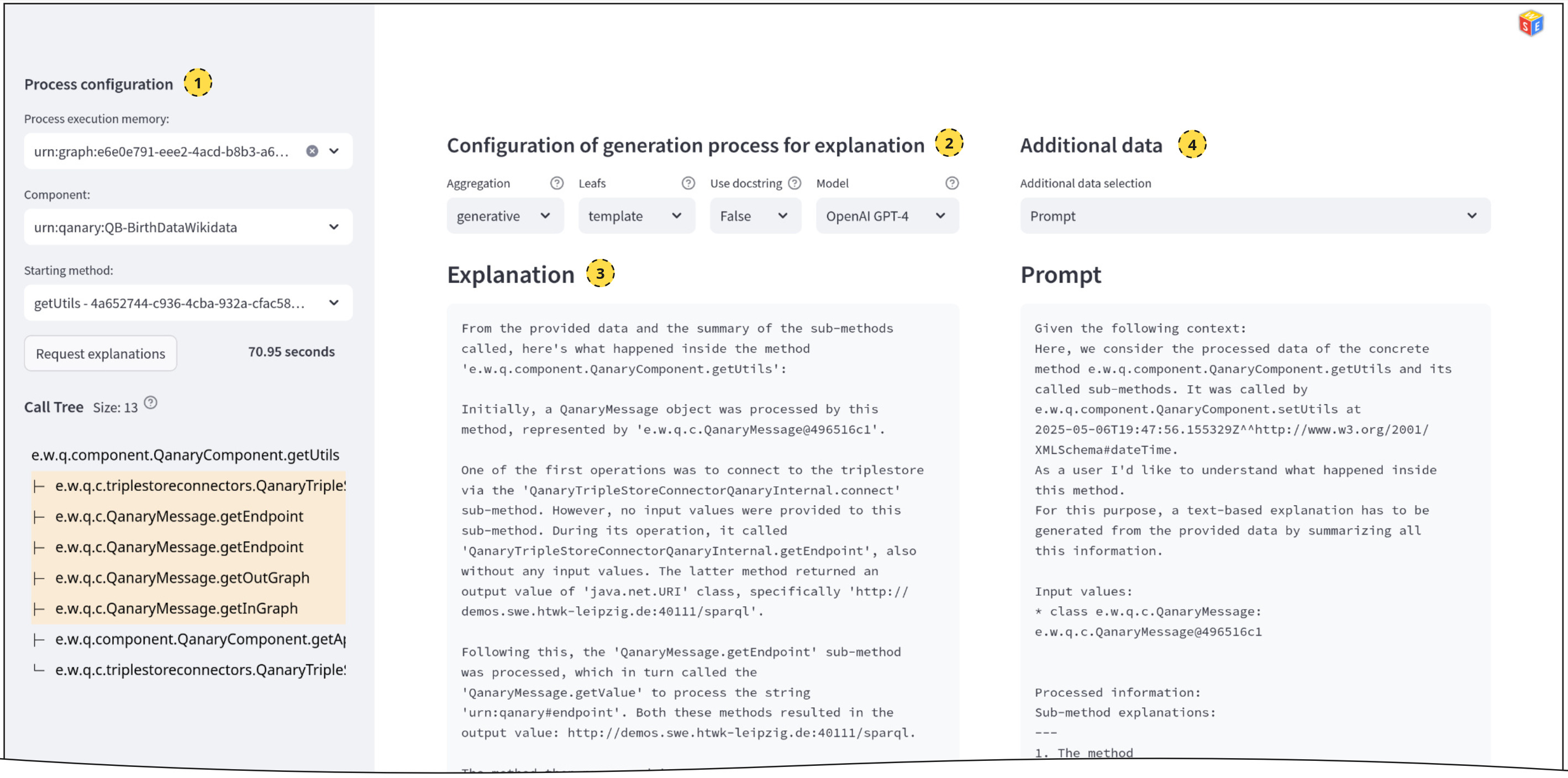} 
    \vspace*{-3ex}
    \caption{Screenshot of the demonstrator. It shows an LLM-generated explanation and the corresponding prompt for the selected process, component, and method of the considered (currently debugged) exemplary component-based system.}
    \label{fig:screenshot}
    \vspace*{-3ex}
\end{figure}

In our approach, we are considering any, distributed or not, (component-based) system, \eg a set of Web services. 
As our approach is built on top of the concrete process data, we established a triplestore for storing the data persistently. 
As an exemplary system, we use Qanary\footnote{\url{https://github.com/WDAqua/Qanary}}. 
It was chosen as its architecture utilizes Web services, which makes it particularly well-suited for analyzing distributed systems — and a special and challenging case for debugging. 
After an executed process, the stored data of each method call (in any component) is utilized in a post-hoc manner to generate the explanations using a Java service\footnote{\url{https://github.com/WSE-research/qanary-explanation-service}}.
We have implemented a template-based generation and one using a user-selected LLM.

\newcommand{\myhigh}[1]{\textit{#1 --}\xspace}
As illustrated in Fig.~\ref{fig:screenshot}, the demonstrator is a frontend that facilitates the explanation of concrete methods for previously executed calls to the considered exemplary (but real-world) component-based system. 
It is available online on our research group's website\footnote{\url{https://wse-research.org/llm-driven-debugging/}} and GitHub\footnote{MIT License licensed source code is available at \\ \url{https://github.com/WSE-research/frontend_aggregated_explanations}}. 
A video walkthrough is available\footnote{\url{https://wse-research.org/LLM-supported-debugging-video}}.

\myhigh{Process configuration \rectangled{1}} Users select one out of the last (already executed) processes, the component where at least one method creates a log entry, and the start method (\ie from the entire call tree of a process, the root method of the subtree, which should be explained). 
Following the vision of providing a complete explanation of the behavior of the application, it is imperative that all subordinate method calls (\ie those from the subtree) are explained. 
Moreover, these explanations are utilized when generating hierarchically higher-level method explanations, either by incorporating them within the prompt or the template. 
Methods from the call sequence (also shown subsequently called methods from any component) can be selected to generate the explanation and show it.

\myhigh{Configuration of generation process for explanations \rectangled{2}} 
Users can decide on how to generate the explanation for the selected method from the call tree. 
As indicated previously, subordinate calls are also explained on demand. 
Two scenarios are defined: First, a method that calls no others (leaf of the call tree) is explained using only the method's process data. 
Second, caller methods aggregate the explanations of their subordinate calls by summarizing them, while also including their own processed data. 
Explanations -- both leaf and aggregate -- can be generated using templates or the selected LLMs. 
The user can also choose whether to include a method's docstring and which LLM to use. 

\myhigh{Explanation \rectangled{3}} The explanation of the selected method is shown. 
Here, users can test which configuration returns the best explanations for their needs. 

\myhigh{Additional data \rectangled{4}} Users can select and show several other data, such as the prompt for generative-based explanations or the used docstring - if applicable.


\section{Conclusions and Future Work}

In this paper, we presented an approach for improved and efficient manual debugging of an exemplary software system. 
The idea of storing actual process data (input and output) during the execution of the application and using this data to explain executed processes appears applicable to systems of any scale. 
In particular, our approach is language-agnostic and is suited for explaining distributed component systems.
The demonstrator shows the usefulness of this data by generating explanations of various methods, processes, and components through LLMs so that an efficient analysis of the software system under consideration is possible.
Our primary focus of this demonstrator was to show the feasibility of our approach built on top of process-dependent data such as input arguments and return values. 
However, by incorporating docstrings, we also took an initial step toward using supporting (static) data, which may lead to further quality improvements. 
In the future, further additional data, such as the source code or performance information (if available), could further improve quality.

\bibliographystyle{splncs04}
\bibliography{literature}

\end{document}